\documentclass[nofootinbib,twocolumn,preprintnumbers]{revtex4-1}
\pdfoutput=1
\usepackage{amsmath,amssymb,graphicx,booktabs,bm,psfrag,color}

\begin{document}

\title{\boldmath Analyzing the CP Nature of a New Scalar Particle via $S\to Zh$ Decay}

\preprint{MITP/16-105}
\preprint{September 30, 2016}

\author{Martin Bauer$^a$}
\author{Matthias Neubert$^{b,c}$}
\author{Andrea Thamm$^b$}

\affiliation{$^a$Institut f\"ur Theoretische Physik, Universit\"at Heidelberg, Philosophenweg 16, 69120 Heidelberg, Germany\\
${}^b$PRISMA Cluster of Excellence {\em\&} MITP, Johannes Gutenberg University, 55099 Mainz, Germany\\
${}^c$Department of Physics {\em\&} LEPP, Cornell University, Ithaca, NY 14853, U.S.A.}

\begin{abstract}
Scalar particles $S$ which are singlets under the Standard Model gauge group are generic features of many models of fundamental physics, in particular as possible mediators to a hidden sector. We show that the decay $S\to Zh$ provides a powerful probe of the CP nature of the scalar, because it is allowed only if $S$ has CP-odd interactions. We perform a model-independent analysis of this decay using an effective Lagrangian and compute the relevant Wilson coefficients arising from integrating out heavy fermions to one-loop order. 
\end{abstract}

\maketitle

\section{Introduction}

Pseudoscalar singlets play an important role in various extensions of the Standard Model (SM). They appear, e.g., as mediators to a dark sector or in solutions to the strong CP problem. Searches at the LHC focus on the model-specific signals of these new states, which often do not reveal their pseudoscalar nature -- the phantom digamma excess seen in the first 13\,TeV data \cite{Aaboud:2016tru,Khachatryan:2016hje} could have been an example of such a signal. Identifying the CP properties of such a new state will be one of the top priorities if a signal is seen in future data.

Let us consider a new spin-0 particle $S$, which is a gauge singlet under the SM gauge group. Assuming its mass is much larger than the electroweak scale, its interactions can be described in terms of local operators in the unbroken phase of the electroweak gauge symmetry. At the renormalizable level, the only interactions of $S$ with SM particles arise from the Higgs portals
\begin{equation}
   {\cal L}_{\rm portal} 
   = - \lambda_1\,S\,\phi^\dagger\phi - \frac{\lambda_2}{2}\,S^2\,\phi^\dagger\phi \,,
\end{equation}
where $\phi$ is the Higgs doublet. The first term gives rise to a mixing between $S$ and the Higgs boson, with a mixing angle $\alpha\sim v\lambda_1/m_S^2$. The coupling $\lambda_1$ is naturally of order the UV cutoff of the theory, but at least of order $m_S$, and hence one expects $\alpha>v/m_S$. However, this mixing will affect the phenomenology of Higgs decay rates, and so in practice $\alpha$ must be small. The example of the elusive 750\,GeV diphoton resonance \cite{Aaboud:2016tru,Khachatryan:2016hje} has demonstrated that tight bounds on $\alpha$ can also be derived from the decays $S\to ZZ$, $WW$, $t\bar t$, $hh$ \cite{Bauer:2016lbe,Dawson:2016ugw}. The portal coupling $\lambda_2$, on the other hand, does not give rise to dangerous effects. 

It is therefore a challenge to model building to find ways of suppressing the coupling $\lambda_1$, either by means of a symmetry or dynamically. A discrete $Z_2$ symmetry under which $S$ changes sign would enforce $\lambda_1=0$. If the ultraviolet theory is (at least approximately) CP invariant, then neutral particles can be classified as CP eigenstates. If $S$ is a CP-odd pseudoscalar ($J^{PC}=0^{-+}$), $\lambda_1$ must be zero. A nice example of a dynamical suppression is provided by models in which $S$ is identified with the lowest mode of a $Z_2$-odd bulk scalar in a warped extra dimension \cite{Bauer:2016lbe,Csaki:2016kqr}. When the Higgs sector is localized on the IR brane, its coupling to $S$ is either suppressed by a small wave-function overlap or by a loop factor. Here we entertain the possibility of eliminating the portal coupling $\lambda_1$ by supposing that $S$ is a CP-odd pseudoscalar, e.g.\ an axion-like particle.

Measurements of angular distributions in $S\to ZZ\to 4l$ or $S\to Z\gamma\to 4l$ decays have been considered as a way of probing the spin and CP properties of a new resonance \cite{Chala:2016mdz,Franceschini:2016gxv}, in analogy with the corresponding measurements in Higgs decays \cite{Soni:1993jc}. However, the rates for these decays are likely to be quite small, since a gauge-singlet $S$ has no renormalizable couplings to gauge bosons. Hence it may require very large statistics to perform these analyses. In this Letter we propose the decay $S\to Zh$, which is strictly forbidden for a CP-even scalar, as a novel and independent way to test the spin and CP quantum numbers of a new particle $S$. The very existence of this decay would constitute a smoking-gun signal for a pseudoscalar nature of $S$ (or for significant CP-odd couplings, in case $S$ is a state with mixed CP quantum numbers), without the need to analyze angular distributions. The observation of this decay would also exclude a spin-2 explanation of a hypothetical new resonance \cite{Kim:2015vba}. To the best of our knowledge this signature has not been studied in the literature. Established experimental searches in the context of two-Higgs-doublet models can be adapted for the proposed search. The most promising decay mode is $S\to Zh\to l^+ l^- b\bar b$ \cite{ATLAS-CONF-2016-015}.

\section{Effective Lagrangian Analysis}
\label{sec:pheno}

At the level of dimension-5 operators, the most general couplings of a CP-odd scalar to gauge bosons read
\begin{equation}\label{SGG}
   {\cal L}_{\rm eff}^{\rm gauge}
   = \frac{\tilde c_{gg}}{M}\,\frac{\alpha_s}{4\pi}\,S\,G_{\mu\nu}^a \widetilde G^{\mu\nu,a}
    + \dots \,,
\end{equation}
where $M$ denotes the new-physics scale, and the dots represent analogous couplings to the $SU(2)_L$ and $U(1)_Y$ gauge bosons. Via this operator the resonance $S$ can be produced in gluon fusion at the LHC. The most general dimension-5 couplings of $S$ to fermions have the same form as the SM Yukawa interactions times $S/M$, and with the Yukawa matrices replaced by some new matrices. In any realistic model these couplings must have a hierarchical structure in the mass basis in order to be consistent with the strong constraints from flavor physics \cite{Goertz:2015nkp}. It is thus reasonable to assume that the dominant couplings are those to the top quarks, see (\ref{Leff5}) below.

When using an effective Lagrangian to describe the production and decays of the resonance $S$ one should keep in mind that in many new-physics scenarios the masses of the heavy particles which are integrated out are in the TeV range. When there is no significant mass gap between $S$ and the new sector, contributions from operators with dimension $D\ge 6$ are not expected to be strongly suppressed. Some of these operators can induce new structures not present at dimension-5 level. 

\subsection{\boldmath $D=5$ operator analysis of $S\to Zh$ decay}
\label{subsec:dim5}

\begin{figure}
\includegraphics[width=0.47\textwidth]{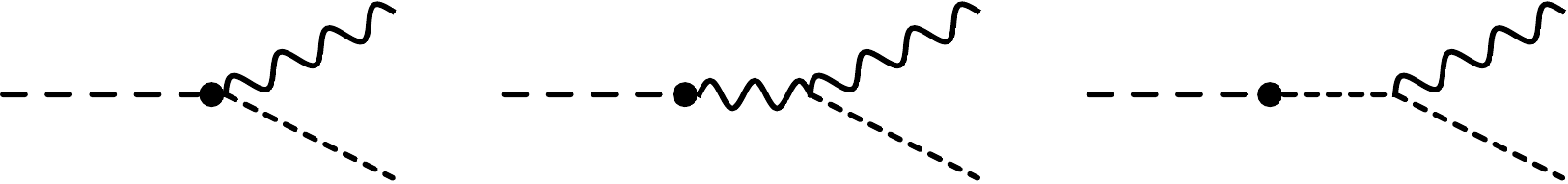}
\caption{\label{fig:EoMgraphs} 
Tree-level diagrams representing the contribution of the operator in (\ref{O5cand}) to $S\to Zh$ decay. The internal dashed line in the third graph represents the Goldstone boson $\varphi_3$.}
\end{figure}  
  
The decay $S\to Zh$ has been studied in the context of two-Higgs-doublet models, where it arises at the renormalizable level via the kinetic terms \cite{Baer:1992uu,Kominis:1994fa}. However, this requires the pseudoscalar $S$ to be light (since the effect vanishes in the decoupling limit) and carry electroweak quantum numbers. In this case the existence of CP-odd couplings of the heavy scalar bosons can be related to three $U(2)$ invariants of the scalar potential \cite{Gunion:2005ja}. For the case of a gauge-singlet scalar considered here no such invariants exist. Moreover, the effective Lagrangian up to dimension~5 does not contain any polynomial operator which could mediate the decay $S\to Zh$ at tree level. The obvious candidate
\begin{equation}\label{O5cand}
   (\partial^\mu S) \left( \phi^\dagger iD_\mu\,\phi + \mbox{h.c.} \right) 
   \to - \frac{g}{2c_w}\,(\partial^\mu S)\,Z_\mu\,(v+h)^2 \,,
\end{equation}
where $c_w\equiv\cos\theta_w$ and the second expression holds in unitary gauge, can be reduced to operators containing fermionic currents using the equations of motion. This follows from the partial conservation of the Higgs current 
\begin{equation}\label{Higgscurrent}
   \vspace{-1.5mm}
   \partial^\mu\!\left( \phi^\dagger iD_\mu\,\phi + \mbox{h.c.} \right)
   \to - \Big( 1 + \frac{h}{v} \Big) \sum_f\,2T_3^f m_f \bar f\,i\gamma_5 f \,,
\end{equation}
where $T_3^f$ is the third component of weak isospin. The resulting operators do not give rise to a tree-level $S\to Zh$ matrix element. Indeed, adding up the diagrams shown in Figure~\ref{fig:EoMgraphs} one finds that the tree-level $S\to Zh$ matrix element of the operator in (\ref{O5cand}) vanishes identically, and the same is true for the $S\to Zhh$ matrix element. 

\begin{figure}
\includegraphics[width=0.47\textwidth]{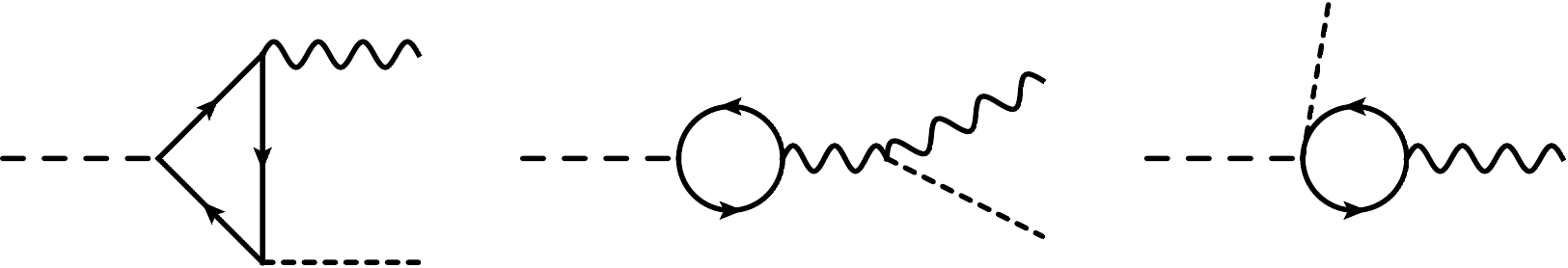}
\caption{\label{fig:toploops} 
Top-loop contributions to $S\to Zh$ decay. We omit a mirror copy of the first graph with a different orientation of the fermion loop and diagrams involving Goldstone bosons.}
\end{figure}

At one-loop order, the $S\to Zh$ decay amplitude receives a contribution from an operator containing quark fields, and since the Higgs boson couples proportional to the quark mass it suffices to consider the term involving the top quark. The relevant Lagrangian is
\begin{equation}\label{Leff5}
   {\cal L}_{\rm eff}^{D=5} 
   = - \tilde c_{tt}\,\frac{y_t}{M}\,S \left( i\bar Q_L\tilde\phi\,t_R + \mbox{h.c.} \right) ,
\end{equation}
where $Q_L$ is the third-generation left-handed quark doublet and $\tilde\phi=\epsilon\phi^*$. The one-loop Feynman diagrams contributing to the decay $S\to Zh$ are shown in Figure~\ref{fig:toploops}. Analogous diagrams involving electroweak gauge bosons in the loop vanish, since it is impossible to saturate the Lorentz indices of the $\epsilon^{\mu\nu\alpha\beta}$ tensor associated with the dual field strength in CP-odd interactions such as (\ref{SGG}). We have evaluated the diagrams in Figure~\ref{fig:toploops} in a general $R_\xi$ gauge. The resulting decay amplitude is
\begin{equation}\label{ourresult}
\begin{aligned}
   i{\cal A}(S\to Zh) &= - \frac{2m_Z\,\epsilon_Z^*\cdot p_h}{M}\,C_5^{\rm top} \,, \\
   \mbox{with} \quad 
   C_5^{\rm top} &= - \frac{N_c\,y_t^2}{8\pi^2}\,T_3^t\,\tilde c_{tt}\,F \,,
\end{aligned}
\end{equation}
where $T_3^t=\frac12$. The $Z$ boson is longitudinally polarized, and hence the structure $2m_Z\,\epsilon_Z^*\cdot p_h\approx 2p_Z\cdot p_h\approx m_S^2$ is proportional to the mass squared of the heavy particle. The quantity $F$ denotes the parameter integral
\begin{equation}\label{Fres}
   F = \int_0^1\!d[xyz]\,\frac{2m_t^2-x m_h^2-z m_Z^2}{m_t^2-xz m_S^2-xy m_h^2-yz m_Z^2-i0} \,,
\end{equation}
with $d[xyz]\equiv dx\,dy\,dz\,\delta(1-x-y-z)$. The factor $y_t^2=2m_t^2/v^2$ in (\ref{ourresult}) ensures that analogous contributions from light fermions in the loop are negligible. Evaluating the integral with $m_t\equiv m_t(m_S)$ and with the physical Higgs and $Z$-boson masses gives $F\approx-0.010+0.673\,i$ for $m_S=750$\,GeV and $F\approx-0.092+ 0.230\,i$ for $m_S=1.5$\,TeV, where here and below we pick two representative values for the mass of the pseudoscalar resonance. For $m_S^2\gg m_t^2$, the function $F$ is formally suppressed by a factor $m_t^2/m_S^2$, but its imaginary part is numerically enhanced. From the amplitude (\ref{ourresult}) we obtain the decay rate
\begin{equation}\label{Gd5}
   \Gamma(S\to Zh)_{D=5} = \frac{m_S^3}{16\pi M^2} \left| C_5^{\rm top} \right|^2
   \lambda^{3/2}(1,x_h,x_Z) \,,
\end{equation}
where $x_i=m_i^2/m_S^2$ and $\lambda(x,y,z)=(x-y-z)^2-4yz$. We find $\Gamma(S\to Zh)_{D=5}\approx 0.6\,\mbox{MeV}\,\tilde c_{tt}^2\,(\mbox{TeV}/M)^2$ in both cases. Assuming that the dominant contribution to the $S\to Zh$ decay amplitude indeed arises at dimension~5, one can derive the model-independent relation
\begin{equation}\label{eq17}
   \frac{\Gamma(S\to Zh)_{D=5}}{\Gamma(S\to t\bar t)}
   = \frac{3y_t^2}{16\pi^2} \left( \frac{m_S}{4\pi v} \right)^2 \! \left| F \right|^2
    \frac{\lambda^{3/2}(1,x_h,x_Z)}{\sqrt{1-4x_t}} \,.
\end{equation}
This ratio evaluates to $3.6\cdot 10^{-4}$ for $m_S=750$\,GeV and $1.8\cdot 10^{-4}$ for $m_S=1.5$\,TeV. The present experimental upper bounds on the corresponding $S\to t\bar t$ rates of about 0.7\,pb and 65\,fb at $\sqrt{s}=8$\,TeV \cite{Aad:2015fna} yield $\sigma(pp\to S_{750}\to t\bar t)<3.2$\,pb and ${\sigma(pp\to S_{1500}\to t\bar t)}<0.6$\,pb at $\sqrt{s}=13$\,TeV under the assumption of gluon-initiated production. Relation (\ref{eq17}) then implies the bounds $\sigma(pp\to S_{750}\to Zh)_{D=5}<1.1$\,fb and $\sigma(pp\to S_{1500}\to Zh)_{D=5}<0.1$\,fb, which are two orders of magnitude below the direct experimental upper limits $\sigma(pp\to S_{750}\to Zh)<123$\,fb and $\sigma(pp\to S_{1500}\to Zh)<40$\,fb at $\sqrt{s}=13$\,TeV \cite{ATLAS-CONF-2016-015}. Note that the former bounds do not apply if $m_S<2m_t$.

\subsection{\boldmath $D=7$ operator analysis of $S\to Zh$ decay}

The dominance of the loop-induced dimension-5 contribution to the $S\to Zh$ decay rate is far from guaranteed. This contribution can be very small if the CP-odd coupling $\tilde c_{tt}$ of $S$ to top quarks is suppressed. Also, as we have seen, the one-loop matrix element in (\ref{ourresult}) is suppressed by a factor $m_t^2/m_S^2$. If $m_S$ is not much smaller than the new-physics scale $M$, the loop contributions arising at dimension~7 can give rise to similar effects. Moreover, at dimension~7 there exists a unique operator giving rise to a tree-level contribution to the $S\to Zh$ amplitude. It reads
\begin{equation}
\begin{aligned}
   O_7 &= (\partial^\mu S) \left( \phi^\dagger iD_\mu\,\phi + \mbox{h.c.} \right)
    \phi^\dagger\phi \\
   &\hspace{1mm}\hat{=}\, - S \left( \phi^\dagger iD_\mu\,\phi + \mbox{h.c.} \right)
    \partial^\mu (\phi^\dagger\phi) \\
   &\to \frac{g}{2c_w}\,S\,Z_\mu\,(v+h)^3\,\partial^\mu h \,,
\end{aligned}
\end{equation}
where in the second step we have used an integration by parts and the equations of motion for the Higgs field, neglecting the fermionic terms in (\ref{Higgscurrent}), which do not contribute to $S\to Zh$ decay at tree level. The expression in the third line, valid in unitary gauge, gives rise to non-vanishing $S\to Zh$ and $S\to Zhh$ matrix elements. 

At one-loop order there exist several dimension-7 operators contributing to the decay $S\to Zh$. Those which mix with $O_7$ under renormalization are
\begin{equation}\label{Leff7}
\begin{aligned}
   {\cal L}_{\rm eff}^{D=7} 
   &= \frac{C_7}{M^3}\,O_7 + \frac{c_6^t}{M^2}\,
    \bar t_R\,\tilde\phi^\dagger i\rlap{\,/}{D}\,\tilde\phi\,t_R \\
   &\quad\mbox{}+ \frac{c_{7a}^t}{M^3}
    \left( i S\,\bar Q_L i\rlap{\,/}{D}\,i\rlap{\,/}{D}\,\tilde\phi\,t_R + \mbox{h.c.} \right) \\
   &\quad\mbox{}+ \frac{c_{7b}^t}{M^3}\,(\partial^\mu S)\,
    \bar t_R\,\tilde\phi^\dagger\gamma^\mu\tilde\phi\,t_R + \dots \,,
\end{aligned}
\end{equation}
plus analogous operators containing the right-handed bottom quark. The dimension-6 operator proportional to $c_6^t$ contributes in conjunction with the operator in (\ref{Leff5}) to give a contribution of order $1/M^3$. 

\begin{figure}
\includegraphics[width=0.47\textwidth]{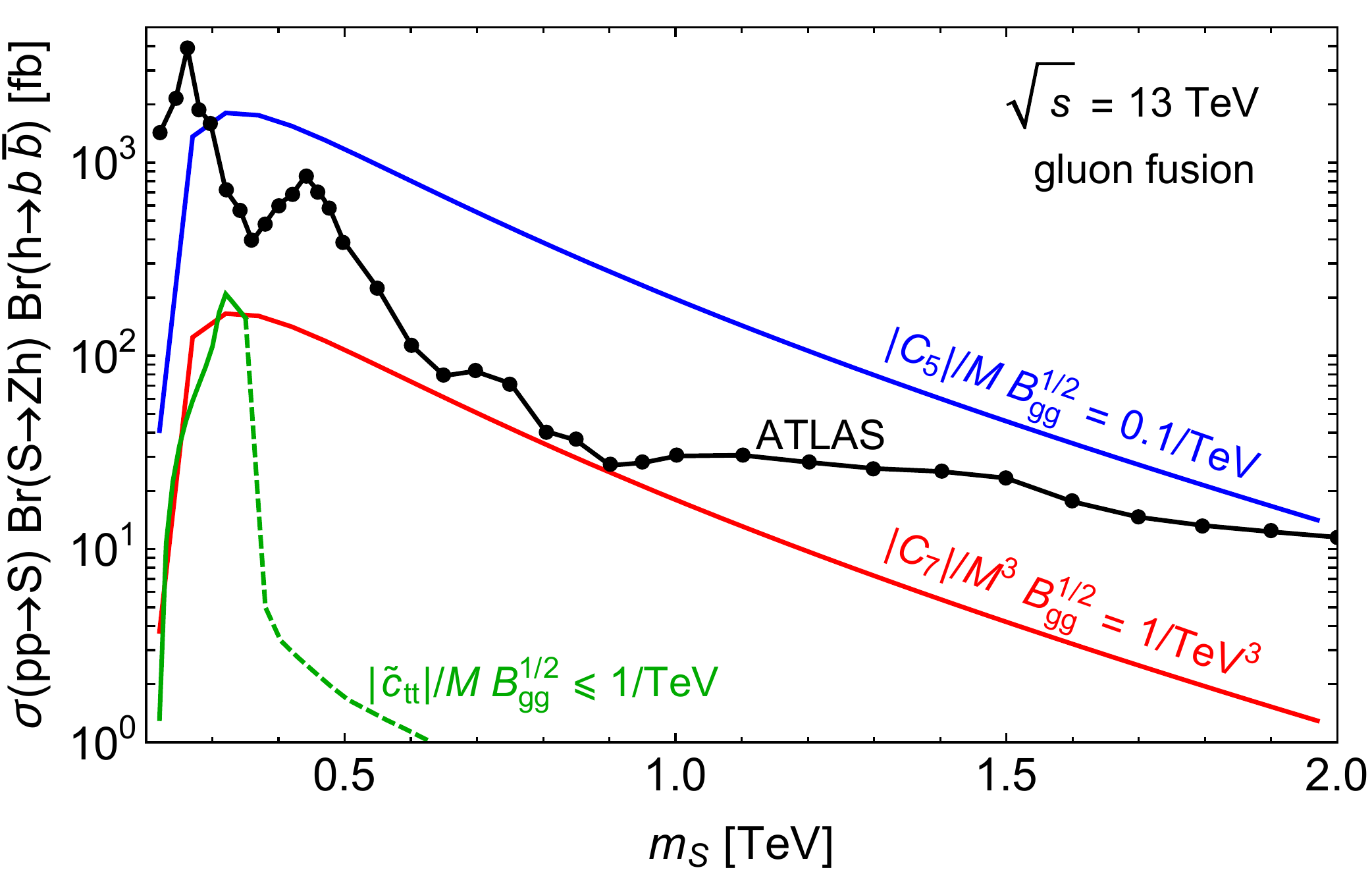}
\caption{\label{fig:new} 
Predictions for the $pp\to S\to Zh\to Zb\bar b$ signal rate vs.\ $m_S$, compared with the ATLAS upper bounds \cite{ATLAS-CONF-2016-015}. The red line shows the contribution from $C_7$ evaluated with $B_{gg}^{1/2}\,|C_7|/M^3=1/\mbox{TeV}^3$, while the blue line shows a generic dimension-5 contribution with $B_{gg}^{1/2}\,|C_5|/M=0.1/\mbox{TeV}$ (see Section~\ref{subsec:poly}), where $B_{gg}\equiv\mbox{Br}(S\to gg)$. The green line shows the contribution from $C_5^{\rm top}$ for $B_{gg}^{1/2}\,|\tilde c_{tt}|/M=1/\mbox{TeV}$, while the dashed green line incorporates the upper bound on $|\tilde c_{tt}|$ implied by the ATLAS limits on the $pp\to S\to t\bar t$ rate \cite{Aad:2015fna}.}
\end{figure}  

\begin{figure}
\includegraphics[width=0.335\textwidth]{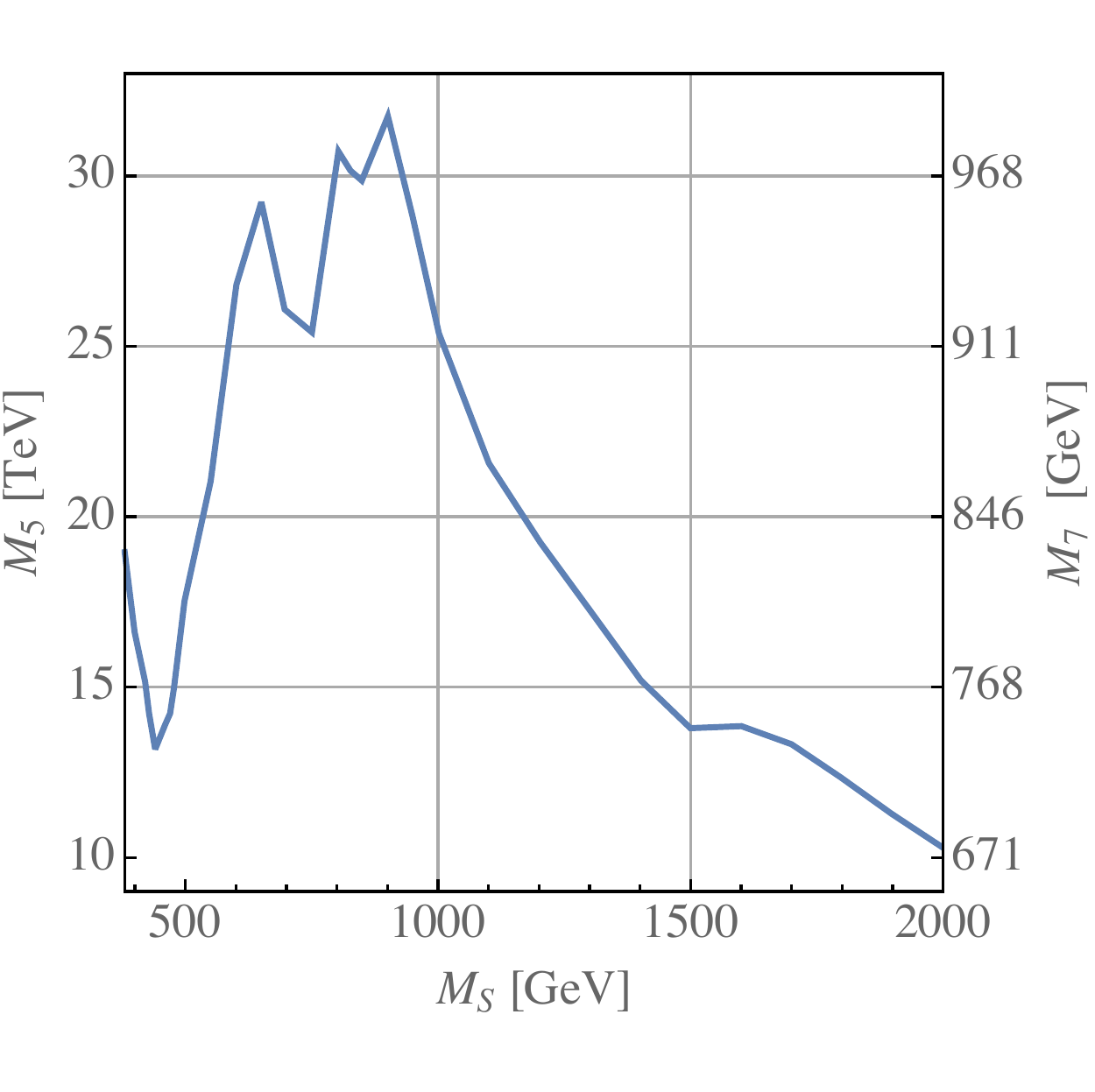}
\vspace{-2mm}
\caption{\label{fig:Mbounds} 
Bounds on the effective new-physics scales $M_5$ and $M_7$ implied by the ATLAS search for $S\to Zh$ decay \cite{ATLAS-CONF-2016-015}.}
\end{figure}

Let us focus on the potentially dominant tree-level contribution from $O_7$, which yields the decay rate  
\begin{equation}\label{GammaSZh7}
   \Gamma(S\to Zh) \approx \frac{m_S^3}{16\pi M^2} \left| C_5^{\rm top} + \frac{v^2}{2M^2}\,C_7 \right|^2
   \lambda^{3/2}(1,x_h,x_Z) \,.
\end{equation}
With $C_7=1$ and $M=1$\,TeV this partial width is about 7\,MeV for $m_S=750$\,GeV and 60\,MeV for $m_S=1.5$\,TeV. The contribution from $C_5^{\rm top}$ can be safely neglected in this case, except in the kinematic region where $m_S<2m_t$. However, in Section~\ref{subsec:poly} below we will consider a more general class of new-physics models, where the coefficient $C_5^{\rm top}$ is replaced by a generic coefficient $C_5$. Figure~\ref{fig:new} shows our results for the $pp\to S\to Zh\to Zb\bar b$ signal rate under the assumption that $S$ is produced in gluon fusion and that a single Wilson coefficient gives the dominant contribution to the $S\to Zh$ rate. The relevant rate can then be written as
\begin{eqnarray}
   &\sigma&(pp\to S)\,\mbox{Br}(S\to Zh)
    = \frac{\pi m_S^2}{128 s}\,\frac{K_{pp\to S}}{K_{S\to gg}}\,\lambda^{3/2}(1,x_h,x_Z) \nonumber\\
   &&\qquad\times f\hspace{-1.5mm}f_{gg}\Big(\frac{m_S^2}{s}\Big)\,\mbox{Br}(S\to gg)\,
    \bigg| \frac{C_5}{M} + \frac{v^2 C_7}{2 M^3} \bigg|^2 ,
\end{eqnarray}
where $f\hspace{-1.5mm}f_{gg}$ is the gluon luminosity function, and the $K_{pp\to S}$ and $K_{S\to gg}$ factors accounting for higher-order QCD corrections have been computed in \cite{Bauer:2016lbe}. We fix the products $B_{gg}^{1/2}\,|C_5|/M$, $B_{gg}^{1/2}\,|C_7|/M^3$ and $B_{gg}^{1/2}\,|\tilde c_{tt}|/M$ (for the case of the top-quark contribution $C_5^{\rm top}$) to the values shown in the plot, denoting $B_{gg}\equiv\mbox{Br}(S\to gg)$. The rate scales with the squares of these combinations. Our results show that $S\to Zh$ rates close to the present experimental bounds are possible for reasonable parameter values, provided that the $S\to gg$ branching ratio is not too small. They can be translated into lower bounds on the effective new-physics scales
\begin{equation}\label{M5M7def}
   M_5\equiv\frac{M}{\left|C_5\right| B_{gg}^{1/2}} \,, \qquad
   M_7\equiv\frac{M}{\left|C_7\right|^{1/3} B_{gg}^{1/6}} \,,
\end{equation}
which are probed by the ATLAS analysis in \cite{ATLAS-CONF-2016-015}. These bounds are shown in Figure~\ref{fig:Mbounds}. They nicely illustrate the new-physics reach of present $S\to Zh$ searches. 

\subsection{\boldmath Non-polynomial operators}
\label{subsec:poly}

It is interesting to consider the hypothetical limit $m_t\gg m_S$ in (\ref{Fres}). Then the parameter integral yields $F=1+{\cal O}(m_S^2/m_t^2)$. The fermion is a very heavy particle, which can be integrated out from the low-energy theory. The contribution (\ref{ourresult}) then corresponds to a one-loop matching contribution to the Wilson coefficient of a local dimension-5 operator with a tree-level $S\to Zh$ matrix element. Our operator analysis in Section~\ref{subsec:dim5} did not reveal the existence of such an operator. However, in extensions of the SM containing heavy particles whose masses arise (or receive their dominant contributions) from electroweak symmetry breaking, operators with a non-polynomial dependence on the Higgs field can arise \cite{Pierce:2006dh}. The non-polynomial structure appears because the particle integrated out (the hypothetical heavy fermion) receives its mass from electroweak symmetry breaking, so it is heavy only in the broken phase of the theory. In our case, the relevant operator reads
\begin{equation}\label{Obeauty}
\begin{aligned}
   O_5 &= (\partial^\mu S) \left( \phi^\dagger iD_\mu\,\phi + \mbox{h.c.} \right)
    \ln\frac{\phi^\dagger\phi}{\mu^2} \\
   &\hspace{1mm}\hat{=}\, - S \left( \phi^\dagger iD_\mu\,\phi + \mbox{h.c.} \right)
    \frac{\partial^\mu (\phi^\dagger\phi)}{\phi^\dagger\phi} \,,
\end{aligned}
\end{equation}
where in the second step we have again used an integration by parts and neglected fermionic currents. The latter expression has a one-to-one map onto the structure of the parameter integral (\ref{Fres}). 

Consider, as an illustration, a sequential fourth generation of heavy leptons, and assume that the heavy charged state $L$ has a mass $m_L\gtrsim m_S/2$ and a coupling $\tilde c_{LL}$ to the pseudoscalar resonance defined in analogy to (\ref{Leff5}). Integrating out this heavy lepton generates the contribution
\begin{equation}
   C_5 = \frac{y_L^2\,\tilde c_{LL}}{16\pi^2} 
   = \frac{m_L^2\,\tilde c_{LL}}{8\pi^2 v^2} 
   \gtrsim \frac{m_S^2\,\tilde c_{LL}}{32\pi^2 v^2}
\end{equation}
to the Wilson coefficient $C_5/M$ of the operator $O_5$. Using this expression instead of $C_5^{\rm top}$ in (\ref{Gd5}), we obtain the upper bounds $|\tilde c_{LL}|<1.3\,(M/{\rm TeV})$ for $m_S=750$\,GeV and $|\tilde c_{LL}|<0.6\,(M/{\rm TeV})$ for $m_S=1.5$\,TeV. In such a model it would be natural to obtain $S\to Zh$ decay rates close to the present experimental upper bounds, see Figure~\ref{fig:new}.

The effective Lagrangian ${\cal L}_{\rm eff}=(C_5/M)\,O_5$ yields a loop correction to the $T$ parameter given by $\alpha(m_Z)\,T=-\Pi_{ZZ}(0)/m_Z^2\approx C_5^2/(4\pi)^2$. Electroweak precision measurements then imply $|C_5|<0.66$ at 95\% confidence level \cite{Baak:2014ora}. This constraint is much weaker than the bounds derived from $S\to Zh$ decay.

The operator $O_5$ and analogous non-polynomial operators of higher dimension are absent in models where the new heavy particles have masses not related to the electroweak scale. We now study such a model in detail.

\section{Heavy vector-like fermions}

It is instructive to consider a concrete new-physics model, which generates the effective interactions of the scalar resonance with SM particles via loop diagrams involving heavy vector-like fermions that are mixed with the SM fermions. Such a scenario is realized, e.g., in models of partial compositeness or warped extra dimension \cite{Kaplan:1991dc,Grossman:1999ra,Gherghetta:2000qt}. We consider an $SU(2)_L$ doublet $\psi=(T~B)^T$ of vector-like quarks with hypercharge $Y_\psi=\frac16$, which mixes with the third-generation quark doublet of the SM. The most general Lagrangian reads
\begin{align}
   {\cal L} &= \bar\psi\left( i\rlap{\,/}{D} - M \right) \psi + \bar Q_L\,i\rlap{\,/}{D}\,Q_L
    + \bar t_R\,i\rlap{\,/}{D}\,t_R + \bar b_R\,i\rlap{\,/}{D}\,b_R \nonumber\\
   &\quad\mbox{}- y_t \big( \bar Q_L\tilde\phi\,t_R + \mbox{h.c.} \big)
    - \big( g_t \bar\psi\,\tilde\phi\,t_R + g_b \bar\psi\,\phi\,b_R + \mbox{h.c.} \big) \nonumber\\
   &\quad\mbox{}- c_1 S\,\bar\psi\,i\gamma_5\,\psi 
    - i c_2 S \big( \bar Q_L\psi - \bar\psi\,Q_L \big) \,,
\end{align}
where we neglect the small Yukawa coupling $|y_b|\ll 1$ of the bottom quark. The terms in the last line contain the couplings to the pseudoscalar resonance $S$. The mass mixing induced by the couplings $g_i$ leads to modifications of the masses and Yukawa couplings of the SM top and bottom quarks by small amounts of order $g_i^2\,v^2/M^2$. Likewise, the masses of the heavy $T$ and $B$ quarks are split by a small amount $M_T-M_B\approx (g_t^2-g_b^2)\,v^2/(4M)$. 

Integrating out the heavy fermion doublet at tree level, by solving its equations of motion, we generate the operators in the effective Lagrangians (\ref{Leff5}) and (\ref{Leff7}) with coefficients $\tilde c_{tt}=-c_2\,g_t/y_t$ and (for $f=t,b$)
\begin{equation}\label{c6c7res}
   c_6^f = g_f^2 \,, \qquad c_{7a}^f = c_2\,g_f \,, \qquad c_{7b}^f = c_1\,g_f^2 \,.
\end{equation}
The coefficient $c_6^b$ is constrained by precision measurements of the $Z$-boson couplings to fermions performed at LEP and SLD. A recent global analysis finds \cite{Efrati:2015eaa}
\begin{equation}\label{gb2}
   c_6^b = g_b^2 = (0.76\pm 0.27) \left( \frac{M}{\rm TeV} \right)^2 ,
\end{equation}
where the pull away from zero is largely driven by the $b$-quark forward-backward asymmetry $A_b^{\rm FB}$, whose experimental value is about 2.8$\sigma$ smaller than the SM prediction \cite{ALEPH:2005ab}. Our model can resolve this anomaly in a natural way. It is likely that the coupling $g_t$ is at least as large as $g_b$, perhaps even significantly larger. In our model the relation $\tilde c_{bb}/\tilde c_{tt}=(g_b/g_t)\,(m_t/m_b)$ holds, and hence the coupling of the resonance $S$ to bottom quarks defined in analogy with (\ref{Leff5}) can be rather large.

The coefficient $C_7$ in (\ref{Leff7}) is induced at one-loop order by diagrams such as those shown in Figure~\ref{fig:toploops}, where now both heavy and light quarks can propagate in the loops. In order to calculate $C_7$ a proper matching onto the low-energy theory must be performed. We obtain 
\begin{widetext}
\begin{equation}\label{C7result}
\begin{aligned}
   \frac{v^2}{2}\,C_7 &= c_1 \sum_{f=t,b}\,\frac{N_c\,g_f^2}{16\pi^2}\,\bigg\{
    2 T_3^f\,\bigg[ m_f^2 \left( L - \frac32 \right) - \frac{m_h^2}{12} + \frac{m_Z^2}{36} 
    + \frac{g_f^2\,v^2}{4} \bigg] - \frac23\,Q_f s_w^2 m_Z^2 \left( L - \frac32 \right)\! \bigg\} \\
   &\quad\mbox{}+ \tilde c_{tt}\,\frac{N_c\,y_t^2}{16\pi^2}\,\bigg\{ 
    2 T_3^t \left[ 3m_t^2 \left( L - \frac32 \right) - \frac{m_h^2}{2}\,\Big( L-\frac76 \Big) 
    - \frac{m_Z^2}{6} \left( L + \frac{19}{6} \right) - g_t^2 v^2 \left( L - \frac94 \right) \right] 
    + Q_t s_w^2 m_Z^2 \bigg\} \,,
\end{aligned}
\end{equation}
\end{widetext}
where $L=\ln(M^2/\mu^2)$. Note the absence of terms proportional to $m_S^2$ on the right-hand side of this expression, which is a consequence of the fact that there is no corresponding dimension-7 operator. There is a non-trivial operator mixing, such that the scale dependence of the coefficient $C_7$ cancels against the scale dependence of the one-loop matrix elements of the fermionic operators in the effective Lagrangian (\ref{Leff7}), see \cite{Bauer:2016ydr} for details. To estimate the dimension-7 contribution we set $\mu=m_Z$ in (\ref{C7result}) and neglect the fermion-loop contributions in the low-energy theory. All large logarithms $L\approx 4.8$ are included in the Wilson coefficient $C_7$, for which we obtain, assuming $M\approx 1$\,TeV in the argument of the logarithms, 
\begin{equation}
\begin{aligned}
   C_7 &\approx \Big[ c_1 \left( 5.30\,g_t^2 + 0.95\,g_t^4 + 0.16\,g_b^2 - 0.95\,g_b^4 \right) \\
   &\hspace{5.5mm}\mbox{}+ \tilde c_{tt} \left( 10.18 - 6.90\,g_t^2 \right) \!\Big] \cdot 10^{-2} \,.
\end{aligned}
\end{equation}
For natural values of the couplings this coefficient can be rather large. For example, with $g_t=2$ and $g_b=0.87$ set by (\ref{gb2}) we get $C_7\approx(0.36 \,c_1-0.17\,\tilde c_{tt})$. For a 750\,GeV resonance produced in gluon fusion (and dominantly decaying to dijets) the production rate
\begin{equation}
   \sigma(pp\to S)\,\mbox{Br}(S\to Zh)\approx 70\,\mbox{fb} \left( \frac{\rm TeV}{M} \right)^6 C_7^2
\end{equation}
can be sizeable and, especially for $M<1$\,TeV, even come close to the current upper bound of 123\,fb \cite{ATLAS-CONF-2016-015}. For $m_S=1.5$\,TeV the rate is smaller by about a factor 10.

\section{Conclusions}

We have presented the first detailed analysis of the decay $S\to Zh$ of a gauge-singlet, heavy spin-0 particle $S$ and pointed out that this process is allowed only if $S$ has CP-odd interactions. Such a pseudoscalar boson arises in many well-motivated extensions of the SM, including models containing Higgs-portal mediators to a hidden sector and scenarios addressing the strong CP problem. Alternative ways to determine the CP nature of a new boson rely on high statistics to perform analyses of angular distributions, whereas the mere observation of the decay $S\to Zh$ proposed here would establish the presence of a CP-odd coupling.

Using a model-independent analysis based on an effective Lagrangian, we have shown that the decay amplitude receives fermion-loop contributions starting at dimension~5, while tree-level contributions can first arise at dimension~7. In new-physics models containing heavy particles whose masses arise from electroweak symmetry breaking there also exists a non-polynomial dimension-5 operator with a tree-level $S\to Zh$ matrix element. 

We have derived explicit expressions for the relevant Wilson coefficients at one-loop order in a model containing heavy vector-like fermions with CP-odd couplings to $S$, finding that appreciable $pp\to S\to Zh$ production rates, even close to the present experimental bounds, can be obtained for reasonable values of parameters. This motivates a vigorous experimental program to search for $S\to Zh$ decays in the high-luminosity LHC run.

\begin{acknowledgments}
M.B.\ acknowledges the support of the Alexander von Humboldt Foundation. The work of M.N.\ and A.T.\ is supported by the Advanced Grant EFT4LHC of the European Research Council (ERC), the DFG Cluster of Excellence PRISMA (EXC 1098) and grant 05H12UME of the German Federal Ministry for Education and Research (BMBF). 
\end{acknowledgments}


\begin{thebibliography}{99}

\bibitem{Aaboud:2016tru} 
  M.~Aaboud {\it et al.} [ATLAS Collaboration],
  JHEP {\bf 1609}, 001 (2016)
  [arXiv:1606.03833 [hep-ex]].
    
\bibitem{Khachatryan:2016hje} 
  V.~Khachatryan {\it et al.} [CMS Collaboration],
  Phys.\ Rev.\ Lett.\  {\bf 117}, no. 5, 051802 (2016)
  [arXiv:1606.04093 [hep-ex]].

\bibitem{Bauer:2016lbe} 
  M.~Bauer, C.~H\"orner and M.~Neubert,
  JHEP {\bf 1607}, 094 (2016)
  [arXiv:1603.05978 [hep-ph]].
  
\bibitem{Dawson:2016ugw} 
  S.~Dawson and I.~M.~Lewis,
  arXiv:1605.04944 [hep-ph].

\bibitem{Csaki:2016kqr} 
  C.~Csaki and L.~Randall,
  JHEP {\bf 1607}, 061 (2016)
  [arXiv:1603.07303 [hep-ph]].
   
\bibitem{Chala:2016mdz} 
  M.~Chala, C.~Grojean, M.~Riembau and T.~Vantalon,
  Phys.\ Lett.\ B {\bf 760}, 220 (2016)
  [arXiv:1604.02029 [hep-ph]].
  
\bibitem{Franceschini:2016gxv} 
  R.~Franceschini, G.~F.~Giudice, J.~F.~Kamenik, M.~McCullough, F.~Riva, A.~Strumia and R.~Torre,
  JHEP {\bf 1607}, 150 (2016)
  [arXiv:1604.06446 [hep-ph]].
  
\bibitem{Soni:1993jc} 
  A.~Soni and R.~M.~Xu,
  Phys.\ Rev.\ D {\bf 48}, 5259 (1993)
  [hep-ph/9301225].
  
\bibitem{Kim:2015vba} 
  H.~M.~Lee, D.~Kim, K.~Kong and S.~C.~Park,
  JHEP {\bf 1511}, 150 (2015)
  [arXiv:1507.06312 [hep-ph]].

\bibitem{ATLAS-CONF-2016-015} 
  ATLAS Collaboration,
  ATLAS-CONF-2016-015.

\bibitem{Goertz:2015nkp} 
  F.~Goertz, J.~F.~Kamenik, A.~Katz and M.~Nardecchia,
  JHEP {\bf 1605}, 187 (2016)
  [arXiv:1512.08500 [hep-ph]].
  
\bibitem{Baer:1992uu} 
  H.~Baer, C.~Kao and X.~Tata,
  Phys.\ Lett.\ B {\bf 303}, 284 (1993).
  
\bibitem{Kominis:1994fa} 
  D.~Kominis,
  Nucl.\ Phys.\ B {\bf 427}, 575 (1994)
  [hep-ph/9402339].

\bibitem{Gunion:2005ja} 
  J.~F.~Gunion and H.~E.~Haber,
  Phys.\ Rev.\ D {\bf 72}, 095002 (2005)
  [hep-ph/0506227].
  
\bibitem{Aad:2015fna} 
  G.~Aad {\it et al.} [ATLAS Collaboration],
  JHEP {\bf 1508}, 148 (2015)
  [arXiv:1505.07018 [hep-ex]].

\bibitem{Pierce:2006dh} 
  A.~Pierce, J.~Thaler and L.~T.~Wang,
  JHEP {\bf 0705}, 070 (2007)
  [hep-ph/0609049].

\bibitem{Kaplan:1991dc} 
  D.~B.~Kaplan,
  Nucl.\ Phys.\ B {\bf 365}, 259 (1991).

\bibitem{Grossman:1999ra} 
  Y.~Grossman and M.~Neubert,
  Phys.\ Lett.\ B {\bf 474}, 361 (2000)
  [hep-ph/9912408].
  
\bibitem{Gherghetta:2000qt} 
  T.~Gherghetta and A.~Pomarol,
  Nucl.\ Phys.\ B {\bf 586}, 141 (2000)
  [hep-ph/0003129].

\bibitem{Efrati:2015eaa} 
  A.~Efrati, A.~Falkowski and Y.~Soreq,
  JHEP {\bf 1507}, 018 (2015)
  [arXiv:1503.07872 [hep-ph]].

\bibitem{ALEPH:2005ab}
  S.~Schael {\it et al.} [ALEPH, DELPHI, L3, OPAL, SLD, LEP Electroweak Working Group, SLD Electroweak Group and SLD Heavy Flavour Group Collaborations],
  Phys.\ Rept.\  {\bf 427} (2006) 257
  [hep-ex/0509008].

\bibitem{Baak:2014ora} 
  M.~Baak {\it et al.} [Gfitter Group Collaboration],
  Eur.\ Phys.\ J.\ C {\bf 74}, 3046 (2014)
  [arXiv:1407.3792 [hep-ph]].

\bibitem{Bauer:2016ydr} 
  M.~Bauer, M.~Neubert and A.~Thamm,
  arXiv:1607.01016 [hep-ph].
    
\end{thebibliography}
\end{document}